%% file: paper.tex
\documentclass[]{fairmeta}

\input{math_commands.tex}

\usepackage{hyperref}
\usepackage{url}
\usepackage{multirow}
\usepackage{subcaption}
\usepackage{booktabs}
\usepackage{adjustbox}
\usepackage{enumitem}
\usepackage{wrapfig}
\usepackage{xcolor}
\usepackage{threeparttable}
\usepackage{longtable}
\usepackage{makecell}
\usepackage{array}
\usepackage{bbm}
\usepackage{tablefootnote}
\usepackage{listings}

\usepackage{tikz}
\usetikzlibrary{arrows}
\usetikzlibrary{arrows.meta}
\usetikzlibrary{positioning}
\usetikzlibrary{bayesnet}
\usepackage{pgfplots}
\usepackage{hyperref}
\usepackage{times}
\usepackage{latexsym}
\usepackage{lmodern}

\usepackage{pifont}
\newcommand{\xmark}{\ding{55}}%

\usepackage{tablefootnote}

\title{\LARGE MELD: Mel-Spectrogram-Based Speech Language Modeling with Discrete Latent Variables}

\author[1,*]{Sung-Lin Yeh}
\author[2,*]{Wei Zhou}
\author[3]{Gil Keren}
\author[3]{Duc Le}
\author[3]{Zhong Meng}
\author[1]{Hao Tang}
\author[3]{Jay Mahadeokar}
\author[3]{Ozlem Kalinli}
\author[3]{Alexandre Mourachko}

\affiliation[1]{University of Edinburgh}
\affiliation[2]{Google DeepMind}
\affiliation[3]{Meta Superintelligence Labs}

\contribution[*]{Work done at Meta}

\abstract{
Recent speech language models rely on encoders that are optimized separately from autoregressive models. Since these encoders are unaware of the downstream objectives,
the extracted representations may not be optimal for downstream tasks.
To address this limitation, we introduce a discrete latent variable model on mel spectrograms that jointly optimizes the encoder and the speech language model.
Joint optimization not only brings improvements over codec-based and other mel-spectrogram-based baselines on zero-shot Text-to-Speech (TTS) and Speech-to-Text (STT) tasks,
but also effectively alleviates common issues in autoregressive mel spectrogram modeling, such as prolonged silence generation and word omissions.
Samples are available at the \href{https://anonymous29300909.github.io/anonymous2930-demo/}{demo page}.
}

\date{\today}
\correspondence{\email{sunglin.yeh@ed.ac.uk}}

\begin{document}

\maketitle

\input{sections/intro}
\input{sections/methods}
\input{sections/related}
\input{sections/exp}

\input{sections/conclusion}
\input{sections/limitation}
\input{sections/ethical}

\clearpage
\newpage
\bibliographystyle{assets/plainnat}
\bibliography{custom}

\clearpage
\newpage
\beginappendix

\input{sections/appendix}

\end{document}

%% file: math_commands.tex
\usepackage{amsmath,amsfonts,bm}

\def\eqref#1{(\ref{#1})}

\def\1{\bm{1}}

\DeclareMathAlphabet{\mathsfit}{\encodingdefault}{\sfdefault}{m}{sl}
\SetMathAlphabet{\mathsfit}{bold}{\encodingdefault}{\sfdefault}{bx}{n}



%% file: sections/intro.tex
\section{Introduction}
Recent advancements in speech language modeling typically rely on a decoupled two-stage training strategy. First, a speech codec \citep{defossez2022highfi,kumar2023high} or a variational autoencoder (VAE) is trained to encode speech signals via intermediate representations; second, an autoregressive model is trained on these representations \cite{chen2025neural, defossez2024moshi,turetzky2024continuous,sun2024multimodal}. 
While this two-stage approach allows each component to be independently optimized
and reused across different speech systems, the encoder is typically unaware of the downstream tasks. 
It is therefore difficult for codec models or autoencoders to know what to represent and what can be discarded.
In fact, discretized representations often fail to preserve task-relevant information unless they are jointly optimized with the downstream objectives \citep{yeh2024estimating,onda2025differentiable}. 

To enable joint training, the training targets need to be either the wave samples themselves or acoustic features, such as mel spectrograms.
Mel spectrograms are a compelling choice, owing to their the success in speaker verification \citep{desplanques2020ecapa}, speech recognition \citep{gulati2020conformer} and speech generation \citep{ren2020fastspeech}.
Autoregressive modeling of mel spectrograms, however, has its difficulties. 
Generating speech autoregressively often gets trapped in stretched silence or produces constant artifacts \citep{tu2025enabling,song2025ella,chen2024vall}, 
especially with mel spectrograms \citep{chen2023vector,battenberg2025robust}.

In this work, we introduce a \underline{Mel}-Spectrogram-Based \underline{D}iscrete Latent Language Model (MELD) for autoregressive text-speech modeling,
in which the mel encoder and the autoregressive model are jointly optimized.
Specifically, we extend the generative process to a discrete latent space and a continuous mel spectrogram space. With the discrete latent space, we make use of sampling methods over discrete distributions that have been shown to effectively suppress stretched silence \citep{chen2024vall, chen2023vector, song2025ella}. At the same time, modeling the continuous mel spectrogram space avoids potential information loss that can degrade STT performance.

We evaluate MELD against mel-spectrogram-based models like MELLE \citep{meng-etal-2025-autoregressive}, which proposes a continuous Gaussian sampling space.
We show that such sampling space is limited with MELLE's objective compared to a discrete latent space in MELD. 
Empirically, our reproduction of MELLE frequently produces prolonged silence, while the discrete sampling in MELD effectively suppresses this issue. 
On zero-shot TTS continuation benchmarks, we observe consistent improvements over MELLE and other codec-based baselines, including VALL-E \citep{chen2025neural}.
The advantage of joint optimization is even more pronounced for STT. 
Unlike input representations such as \texttt{dMel} that are discretized independently of the STT objective, MELD is optimized directly from mel spectrograms w.r.t.\ the STT objective, leading to lower word error rates.

Finally, we show that MELD opens new opportunities to improve joint TTS–STT modeling.
Training the two tasks together is known to be challenging \citep{battenberg2025robust,bai2024dmel} for several reasons.
For instance, an STT-only model does not necessarily need to preserve prosodic information that is crucial for zero-shot TTS.
We show that MELD can, by design, learn both STT and TTS tasks within a single autoregressive model.
In particular, the strength of joint optimization over mel spectrograms results in significant improvements in STT compared to \texttt{dMel} \citep{bai2024dmel}.

%% file: sections/methods.tex
\begin{figure*}[t]
\centering
\begin{minipage}{0.4\textwidth}
\centering
\begin{tikzpicture}[->,>={Latex[width=1.4mm,length=2mm]},auto,node distance=3cm, thick,
main node/.style={circle, draw=black, fill=blue!17, thick, minimum size=10mm,font=\fontsize{19}{0}\selectfont},
latent/.style={circle, draw=black, fill=gray!0, thick, minimum size=6mm,font=\fontsize{19}{0}\selectfont}, scale=0.55, transform shape]

\node[main node] (x) {$x_{<t}$};
\node[latent] (z) [right of=x] {$z_t$};
\node[main node] (y) [right of=z] {$x_t$};
\node[main node] (text) [above=8mm of x] {$y$};

\path[]
    (x) edge[->,line width=1pt] node [right] {} (z)
    (z) edge[->,line width=1pt] node [right] {} (y)
    (x) edge[->,line width=1pt] [bend right] node [left] {} (y)
    (text) edge[->,line width=1pt] [bend left=15] node [left] {} (z)
    (text) edge[->,line width=1pt] [bend left] node [left] {} (y)
    (y) edge[->,line width=1pt] [dashed, bend right] node [left] {} (z);
\end{tikzpicture}
    \caption{A graphical model of the generative process. We adopt the style of \citet{Kingma2014}, denoting the solid lines as the generative model $p(x_t|z_t,x_{<t},y)p(z_t|x_{<t},y)$, 
    while the dashed line as the proposal distribution $q(z_t|x_t)$.
    }
    \label{fig:graph}
\end{minipage}
\hfill
\begin{minipage}{0.58\textwidth}
\centering
\begin{tikzpicture}[scale=1.15, transform shape, ->,>={Latex[width=1.4mm,length=2mm]},auto,node distance=3cm, thick,
token/.style={circle, draw=black, fill=red!40, thick, minimum size=4.5mm,font=\fontsize{13}{0}\selectfont},
latent/.style={circle, draw=black, fill=orange!80, thick, minimum size=4.5mm,font=\fontsize{13}{0}\selectfont},
eos/.style={circle, draw=black, fill=red!60, thick, minimum size=5.5mm,font=\fontsize{13}{0}\selectfont},
fbank/.style={rectangle, draw=black, fill=blue!40, thick, minimum size=5mm,font=\fontsize{13}{0}\selectfont},
emb_t/.style={rectangle, draw=black, fill=black!20, thick, minimum size=5mm,font=\fontsize{13}{0}\selectfont},
emb_s/.style={rectangle, draw=black, fill=black!20, thick, minimum size=5mm,font=\fontsize{13}{0}\selectfont},
arrow/.style={-latex},
arrowc2/.style={arrow, rounded corners=.5cm},
tts/.style={circle, draw=black, fill=gray, thick, minimum size=5.5mm,font=\fontsize{13}{0}\selectfont},
scale=0.5, transform shape]
\tikzset{arrows={[scale=0.7]}}

\foreach \i in {3,4}
    \node[token] (x\i) at ({(\i-1)*1.25}, -0.7) {};
    
\node[font=\fontsize{15}{0}\selectfont] (x5) at ({(5-1)*1.25}, -0.7) {\Large\texttt{<TTS>}};
\foreach \i in {6,7,8}
    \node[fbank] (x\i) at ({(\i-1)*1.25}, -0.7) {}; 
\foreach \i in {3,4,5}
    \node[emb_t] (e\i) at ({(\i-1)*1.25}, 0.3) {}; 
\foreach \i in {6,7,8}
    \node[emb_s] (e\i) at ({(\i-1)*1.25}, 0.3) {}; 
\foreach \i in {3,4,5,6,7,8}
    \draw[->] (x\i) -- (e\i);
\foreach \i in {3,4,5,6,7,8}
    \draw[->] (e\i) -- ({(\i-1)*1.25}, 1.1);
\node[draw, minimum width=7.2cm, minimum height=1.0cm, fill=gray!0, anchor=south, rounded corners=3pt, thick]
at (5.65, 1.1) {\Large Decoder-only Transformer};

\foreach \i in {5,6,7}
    \node[latent] (z\i) at ({(\i-1)*1.25}, 3.0) {}; 
\foreach \i in {5,6,7}
    \node[fbank] (o\i) at ({(\i-1)*1.25}, 4.1) {}; 
\foreach \i in {5,6,7}
    \draw[->] (z\i) -- (o\i);
    
\foreach \i in {5,6,7}
    \draw[thick, ->, looseness=1] ({(\i-1)*1.25}, 2.1) to[bend left=40] (o\i.south);

\node[font=\fontsize{18}{0}\selectfont] (x10) at ({(8-1)*1.25}, 3.0) {\Large\texttt{<EOS>}};
    
\foreach \i in {5,6,7,8}
    \draw[->] ({(\i-1)*1.25}, 2.1) -- ({(\i-1)*1.25}, 2.8);

\node[token, minimum size=4mm, label=right:\Large BPE token] at ({(0.9+3)*1.25-2.3}, -2.1+0.25) {};

\node[label=right:\Large mel spec] at ({(3.3+3)*1.25-2.1}, -2.15+0.25) {};
\node[fbank,minimum size=4mm] at ({(3.2+3)*1.25-2.1}, -2.1+0.25) {}; 

\node[fbank,minimum size=4mm,fill=black!20,] at ({(5.1+3)*1.25-1.8}, -2.1+0.25) {};
\node[label=right:\Large input embedding] at ({(5.2+3)*1.25-1.8}, -2.15+0.25) {};

\node[latent, minimum size=4mm, label=right:\Large discrete distribution] at ({(8.1+3)*1.25-1.1}, -2.1+0.25) {};

\foreach \i in {3,4,5}
    \node[fbank] (x\i) at ({(\i+5.5)*1.25}, -0.7) {};
\node[font=\fontsize{15}{0}\selectfont] (x6) at ({(6+5.5)*1.25}, -0.7) {\Large\texttt{<STT>}};
\foreach \i in {7,8}
    \node[token] (x\i) at ({(\i+5.5)*1.25}, -0.7) {};

\foreach \i in {3,4,5}
    \node[emb_s] (e\i) at ({(\i+5.5)*1.25}, 0.3) {}; 
\foreach \i in {6,7,8}
    \node[emb_t] (e\i) at ({(\i+5.5)*1.25}, 0.3) {}; 
\foreach \i in {3,4,5,6,7,8}
    \draw[->] (x\i) -- (e\i);
\foreach \i in {3,4,5,6,7,8}
    \draw[->] (e\i) -- ({(\i+5.5)*1.25}, 1.1);
\node[draw, minimum width=7.2cm, minimum height=1.0cm, fill=gray!0, anchor=south, rounded corners=3pt, thick]
at (5.65+8, 1.1) {\Large Decoder-only Transformer};
\foreach \i in {6,7}
    \node[latent] (z\i) at ({(\i+5.5)*1.25}, 3.0) {}; 
\node[font=\fontsize{15}{0}\selectfont] (x10) at ({(8+5.5)*1.25}, 3.0) {\Large\texttt{<EOS>}};
\foreach \i in {6,7,8}
    \draw[->] ({(\i+5.5)*1.25}, 2.1) -- ({(\i+5.5)*1.25}, 2.8);
\end{tikzpicture}
\caption{An illustration of two sequence prediction tasks controlled by two special tokens $\texttt{<TTS>}$ and $\texttt{<STT>}$, respectively. 
}
\label{fig:lm}
\end{minipage}
\end{figure*}

\section{Autoregressive learning objectives}
We briefly review autoregressive modeling of mel spectrograms conditioned on its transcription,
and derive the learning objective of our discrete latent variable model.
Let $y=(y_1,\dots,y_M)$ be a byte-pair-encoding (BPE) token sequence \citep{sennrich2016neural} with each token from a text vocabulary $y_m \in \mathcal{V}_{\text{text}}$,
$x=(x_1,\dots,x_T)$ be a sequence of mel frames where $x_t \in \mathbb{R}^{d_{\text{mel}}}$.
Consider an autoregressive model, the log likelihood of generating acoustic frames given the text factorizes
\begin{align}
-\log p(x|y) = \sum_{t = 1}^{T} -\log p(x_t|x_{<t}, y).
\label{eq:likelihood}
\end{align}
The conditional probability is recently modeled by a decoder-only Transformer. 
Because the model only learns to predict the next frame without knowing when to terminate generation, prior work introduces a jointly trained stop predictor \citep{wang2017tacotron,meng-etal-2025-autoregressive}.
We will show that the stop predictor is unnecessary in our model.

A limitation of such models is that they typically predict deterministic frame estimates and lack a well-defined distribution for sampling.
Generation without sampling tends to get stuck in infinite silence \citep{chen2024vall,song2025ella}, as commonly observed in sequence-to-sequence mel spectrogram TTS models \citep{wang2017tacotron,shen2018natural,chen2023vector}.
There is also no finer control over the output distributions as in discrete sampling \citep{chen2025neural,chen2024vall}.

\subsection{Generation with discrete latent variables}
\label{sec:dlv-lm}

Inspired by the success of discrete sampling in speech language modeling \citep{wang2023neural}, MELD extends the autoregressive process across a discrete latent space and a continuous mel spectrogram space. 
Let $z=(z_1,\dots,z_T)$ be the discrete latent variables with $z_t \in \mathcal{V}_{\text{latent}}$ and $\mathcal{V}_{\text{latent}} = \{1,\dots,K\}$.
The generative factorization over the next Mel frame $x_t$ and latent variable $z_t$ is
\begin{align}
&p(x_t, z_t|x_{<t},y)
= p(x_t|z_t, x_{<t}, y)p(z_t|x_{<t}, y), 
\label{eq:factor}
\end{align}
where $x_{<t}$ and $y$ represent historical frames and text tokens, respectively.

In \eqref{eq:factor}, there are two key conditional independence assumptions. 
First, we assume the past frames $x_{<t}$ and text $y$ provide sufficient information for the next latent $z_t$, making $z_t$ conditionally independent of $z_{<t}$.
Second, given $z_t$ and $x_{<t}$, the next frame $x_t$ is conditionally independent of $z_{<t}$. 
Importantly, conditioning the prediction of $x_t$ solely on $z_t$ is usually insufficient to capture the acoustic details for autoregressive zero-shot TTS, unless one expands the latent space $K$ via residual vector quantization (RVQ) or introduces speaker embeddings. 
Conditioning on $x_{<t}$ and $y$ alongside $z_t$ enables accurate next-frame prediction while maintaining a compact discrete latent space.
We visualize the generative process in Figure~\ref{fig:graph}.

Directly maximizing the marginal log likelihood, i.e., $\log\sum_{z_t=1}^K p(x_t|z_t, x_{<t}, y)p(z_t|x_{<t}, y)$, is computationally intractable for large $K$.
We thus introduce a variational framework to optimize its lower bound in the following section.

\subsection{Variational next-frame prediction}

To efficiently approximate the log likelihood,
we introduce a variational distribution to draw quantized samples from the discrete latent space.
We define $q(z|x,y)=\prod_t q(z_t|x_t)$, assuming frame-wise conditional independence to match the autoregressive generative factorization.
The optimization objective under such assumption can be computed linear in $T$.
The Variational Lower Bound (VLB) is derived as:
\begin{align}
\mathcal{L}_{\text{VLB}} = \sum_{t = 1}^T \Big[
\mathrm{KL}\big[q(z_t|x_t)\,\|\, p(z_t | x_{<t}, y)\big] - \mathbb{E}_{z_t \sim q}\big[\log p(x_t |z_t, x_{<t}, y)\big]
\Big],
\label{eq:vlb}
\end{align}
where the first term is the KL divergence between the frame-wise quantization network $q(z_t|x_t)$ and an autoregressive network $p(z_t|x_{<t}, y)$, the second term is the reconstruction loss of a reconstruction network $p(x_t | z_t, x_{<t}, y)$.
The step-by-step derivation can be found in Appendix~\ref{app:proof}.

\subsection{Extension to speech-to-text}
The connection between our variational objective and standard next-token prediction becomes clear by expanding the KL term in Equation~\eqref{eq:vlb} into cross-entropy and entropy terms:
\begin{align}
\mathbb{E}_{z_t \sim q} &\big[-\log p(z_t | x_{<t}, y) +\log q(z_t|x_t)\big].
\label{eq:ntp-speech}
\end{align}
The cross entropy term over the discrete latent variable $z_t$ allows us to integrate STT into the same sequence modeling framework.
To achieve this,
we extend the target discrete space to a union token set $\mathcal{V} =\mathcal{V}_{\text{text}} \cup \mathcal{V}_{\text{latent}}$, where $\mathcal{V}_{\text{text}}$ represents the BPE tokens and $\mathcal{V}_{\text{latent}} = \{1, \dots, K\}$ represents discrete speech latents.

The STT objective minimizes the cross entropy
\begin{align}
\sum_{t = 1}^{M}\mathbb{E}_{z_t \sim q}\big[-\log p(z_t|y_{<t},x)\big],
\label{eq:ntp-text}
\end{align}
where we sample $z_t$ from $q(z_t|y_t)=\mathbbm{1}_{z_t = y_t}$ and $y=(y_1,\dots,y_M)$ represents BPE tokens.
We model the next-token distributions, $p(z_t|y_{<t}, x)$ for STT with $z_t \in \mathcal{V}_{\text{text}}$, and $p(z_t|x_{<t}, y)$ for TTS with $z_t \in \mathcal{V}_{\text{latent}}$, using the same autoregressive network.

Together, there are in total 3 special tokens as presented in Figure~\ref{fig:lm},
alongside BPE tokens and $K$ discrete latents: \texttt{<TTS>} for speech synthesis conditioning, \texttt{<STT>} for transcribing conditioning,
and \texttt{<EOS>} for the end of sequence. 
\texttt{<EOS>} is merged into the discrete space, and jointly predicted with BPE tokens and discrete latents.

\section{MELD parameterization and training}
The feedforward process of MELD for TTS is visualized in Figure~\ref{fig:feedforward}:
A quantization network $q(z_t|x_t)$ first quantizes the next frame $x_t$. 
Next, an autoregressive network $p(z_t|x_{<t}, y)$ is learned to predict a sample $z_t$ drawn from $q(z_t|x_t)$. Finally, a reconstruction network $p(x_t|z_t, x_{<t}, y)$ reconstructs the next mel frame. We now describe the parameterization of each component step by step.

\subsection{Quantizing the next frame and sampling}
\label{sec:soft-vq}

\begin{figure}
\centering
\begin{tikzpicture}[scale=0.9, transform shape,->,>={Latex[width=1.5mm,length=2mm]},
code/.style={draw, rectangle, fill=green, inner sep=0.05cm,minimum width=0.15cm, minimum height=0.4cm, thick}]

  \node[inner sep=0pt] (q) at (4.32, 2.8) {
    \begin{tikzpicture}
      \begin{axis}[
        ybar,
        symbolic x coords={A,B,C,D},
        xtick=\empty,
        ytick=\empty,
        axis line style={draw=none},
        tick style={draw=none},
        ymin=0, ymax=0.6,
        width=3cm,
        height=3cm,
        bar width=6pt,
        ]
        \addplot+[fill=orange!60,draw=black, thick] coordinates {(A,0.05) (B,0.2) (C,0.3) (D,0.2)};
      \end{axis}
    \end{tikzpicture}
  };
  \node[inner sep=0pt,left=1.2cm of q] (p)  {
    \begin{tikzpicture}
      \begin{axis}[
        ybar,
        symbolic x coords={A,B,C,D},
        xtick=\empty,
        ytick=\empty,
        axis line style={draw=none},
        tick style={draw=none},
        ymin=0, ymax=0.6,
        width=3cm,
        height=3cm,
        bar width=6pt,
        ]
        \addplot+[fill=orange!60,draw=black, thick] coordinates {(A,0.15) (B,0.3) (C,0.2) (D,0.15)};
      \end{axis}
    \end{tikzpicture}
  };
\node[above=-0.6cm of q] (labelq) {\footnotesize$q(z_t|x_t)$};
\node[above=-0.6cm of p] (labelp) {\footnotesize$p(z_t|x_{<t},y)$};

\node[code, label={above:$c_1$}] (v1) at (3.5, 1.05) {};
\node[code, label={above:$c_2$}] (v2) at (3.9, 1.05) {};
\node[code, label={above:$c_3$}] (v3) at (4.3, 1.05) {};
\node (v4) at (4.7, 1.05) {$\cdots$};
\node[code, label={above:$c_K$}] (v5) at (5.1, 1.05) {};
\draw[rounded corners, thick] (4.5-1.27, 0.75) rectangle (4.5+0.85, 1.7
);
  
\node[draw, thick, below=1.8cm of q, fill=blue!0] (xt) {$x_{t}$};
\node[draw, thick, left=2.1cm of xt] (h) {$h_{t}$};
\draw[->, thick] (xt) -- (4.35, 0.75); 
\draw[->, thick] (4.35, 1.7) -- (4.35, 2.1); 

\node[draw, thick, above=0.8cm of h, rounded corners=1mm,rectangle,fill=gray!20,minimum width=1.2cm] (linear) {\small $\mathrm{Linear}$};
\node[draw, thick, rounded corners=1mm,rectangle,fill=gray!20,minimum width=0.9cm] (g) at (3, 2.9) {\small $g_{\text{mel}}$};
\node[above=1.2cm of g, fill=yellow!25, rounded corners=1mm] (text1) {\textbf{\texttt{Latent Prediction}}};
\node[draw, thick, right=0.7cm of v5, rounded corners=1mm,rectangle,fill=gray!20,minimum width=1.2cm] (linear2) at (5.3, 0.7) {\small $\mathrm{Linear}$};

\node[draw, thick, rounded corners=1mm,rectangle,fill=gray!20,minimum width=1cm] (mlp) at (7.8, 1.26
) {\small $\mathrm{MLP}$};
\node[draw, thick, above=0.9cm of linear2, fill=blue!0] (xhat) {$\hat{x}_t$};
\draw[->, thick] (h) -- (linear); 
\draw[->, thick] (1.63, 1.63) -- (1.63, 2.1); 

\node[draw, thick,circle, minimum size=0.4cm, inner sep=0pt, above=1.85cm of linear] (plus) {};
\draw[line width=1pt, -] (plus.center) ++(-0.2cm,0) -- ++(0.4cm,0); 
\draw[line width=1pt, -] (plus.center) ++(0,-0.2cm) -- ++(0,0.4cm); 

\draw[thick, ->, rounded corners=0.1cm] (1.6, 0.6) -- ++(-1,0) -- ++(0,3.1) -- (plus);

\path[]
    (3.5, 2.3) edge [bend left, thick] node [below] {$c_{z_t}$} (g)
    (g.north) edge [bend right, thick] node [left] {} (plus.east)
    (linear2.north) edge [thick] node [left] {} (xhat.south)
    (6.6, 1.18) edge [bend right, thick, looseness=1, in=-140, out=0] node [left] {} (mlp.south)
    (mlp.north) edge [bend right, thick, looseness=1, in=-140, out=-45] node [left] {} (6.6, 1.3);

\draw[thick, ->, rounded corners=0.15cm] (plus.north) -- ++(0,0.2) -- ++(4,0) -- ++(0,-4.2) -- ++(1.04,0) -- ++(0, 0.5);

\node[right=0.1cm of xhat, fill=blue!0] (xt1) {$\hat{x}_{t+1}$};
\node[right=0.1cm of xt1, fill=blue!0] (dr) {$\dots$};
\node[draw, thick, rounded corners=1mm,rectangle,fill=gray!20,minimum width=1cm] (conv) at (8.6, 2.8) {\footnotesize $\mathrm{conv}$};
\node[draw, thick,above=0.85cm of xt1, fill=blue!0] (xout) {$\hat{x}+\mathrm{conv}(\hat{x})$};
\node[right=0.8cm of text1,, fill=yellow!25, rounded corners=1mm] {\textbf{\texttt{Mel Reconstruction}}};

\path[]
    (7.55, 2.7) edge [bend right, thick, looseness=1, in=-140, out=0] node [left] {} (conv.south)
    (conv.north) edge [bend right, thick, looseness=1, in=-140, out=-60] node [left] {} (7.55, 2.86)
    (xt1.north) edge [thick] node [left] {} (xout.south);
\end{tikzpicture}
\caption{Variational autoregressive training process, where $h_t$ summarizes $x_{<t}$ and $y$. 
The model is trained to minimize the KL divergence between $q$ and $p$, and the MSE between $x_t$ and $\hat{x}_t$,
$\hat{x}_t+\mathrm{conv}(\hat{x})_t$.
}
\label{fig:feedforward}
\end{figure}
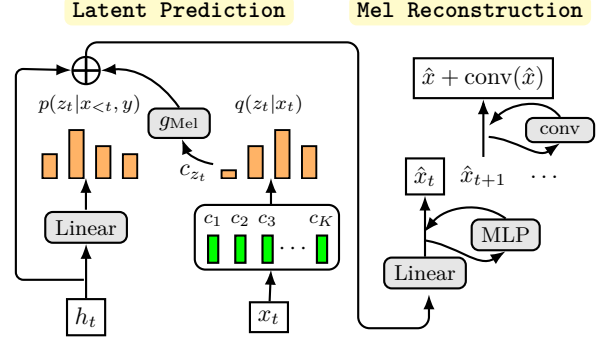

As shown in Figure~\ref{fig:feedforward},
we draw samples from $q(z_t|x_t)$ during training to serve both as targets for computing the KL term and as the latent embeddings for reconstructing mel frames described in Eq~\eqref{eq:vlb}.
To draw a sample that represents $x_t$ well, 
we initialize the codebook with $k$-means centroids trained on mel spectrograms.
The variational distribution assigns a frame based on the Euclidean distance between $x_t$ and each codeword entry,
\begin{align}
q(z_t|x_t) = \frac{\exp\left(-\|x_t - c_{z_t}\|^2 / \tau\right)}
  {\sum_{k=1}^{K} \exp\left(-\|x_t - c_k\|^2 / \tau\right)},
\label{eq:softmax-q}
\end{align}
where a codeword $c_k$ is the $k$-th column of a codebook $C \in \mathbb{R}^{d_{\text{mel}} \times K}$, and $\tau$ is the temperature.
We set $\tau=1$ in this work.

The distribution can be interpreted as the assignment probability under an isotropic Gaussian mixture model (GMM) with an isotropic covariance $\tau I$.
When $\tau \rightarrow 0$, the distribution becomes minimization or hard assignment in $k$-means \citep{murphy2012machine,Kulis2011RevisitingKN}.
This parameterization is also known as \textit{soft} vector quantization \citep{agustsson2017soft,sonderby2017continuous}.

With the codebook being properly initialized, 
each quantized frame represents the original frame well with small distortion.
We freeze the codebook throughout training and let the reconstruction network (Section~\ref{sec:recnet}) refine the codewords.
This avoids potential challenges of training neural networks with vector quantization \citep{huh2023straightening}.
Despite the codebook being fixed, the quantization in our framework does not disconnect gradients from the target mel frames to the input encoder and the autoregressive model, allowing the joint optimization of the encoder and the autoregressive model.
This is a key distinction between MELD and the two-stage approaches \citep{chen2024vall,bai2024dmel}.
In two-stage approaches, an encoder is placed between the wave samples or mel spectrograms and the autoregressive model and held fixed.

The quantization network is only used to model the variational distribution $q$ in the variational bound \eqref{eq:vlb} to learn the generative model \eqref{eq:factor}.
It is therefore not required during inference, when future frames are also not available.

\subsection{Autoregressive latent prediction}
\label{sec:latent-dec}
As shown in Figure~\ref{fig:lm}, we apply a decoder-only Transformer to predict the next discrete latent variable $z_t$ drawn from the variational distribution $q$, or the subsequent text token, i.e., the two prediction distributions $p(z_t|x_{<t},y)$ and $p(z_t|y_{<t}, x)$.
Text tokens and mel spectrograms are mapped to $d_{\text{model}}$-dimensional embeddings before the Transformer.
A lookup function $g_{\text{text}}: \mathcal{V}_{\text{text}}\rightarrow \mathbb{R}^{d_{\text{model}}}$ maps a text token $y_m$ into a text embedding.
We follow \citep{meng-etal-2025-autoregressive}, using an encoder of 3-layer multi-layer perceptron (MLP), $g_{\text{mel}}: \mathbb{R}^{d_{\text{mel}}} \rightarrow \mathbb{R}^{d_{\text{model}}}$ to encode each mel frame.
Figure~\ref{fig:feedforward} (left) presents the feedforward process of latent prediction.

When modeling TTS and STT tasks together, 
the two distributions are modeled by a softmax over the discrete space, the disjoint vocabulary $\mathcal{V} = \mathcal{V}_{\text{text}} \cup \mathcal{V}_{\text{latent}}$, followed by a linear layer.

\subsection{Spectrogram reconstruction}
\label{sec:recnet}
The reconstruction network is similar to that of Tacotron 2 \citep{shen2018natural}, 
consisting of a linear layer with residual connections of a 3-layer multilayer perceptron and a convolutional module. 
The prediction consists of two stages.
In Figure~\ref{fig:feedforward},
an output embedding $h_t$ and a codeword $c_{z_t}$ sampled from $q$ are mapped to a mel spectrogram space in an autoregressive manner. 
An MLP layer is used to predict the residual and added the output of a linear layer to produce $\hat{x}_t$. 
We denote this process as 
\begin{align}
    \hat{x}_t& = \mathrm{SpecNet}(h_t + g_{\text{mel}}(c_{z_t})) \text{ where}, \notag\\
    &z_t \sim q(z_t|x_t) \text{ during training,} \notag\\
    &\text{$z_t \sim p(z_t|x_{<t}, y)$ during inference.}
    \label{eq:specdec}
\end{align}
During inference, we sample $z_t$ from the predicted discrete distributions.
To ensure the discrete signals $z_t$ will not get overlooked, 
we embed the codevector $c_{z_t}$ with $g_{\text{mel}}$ that is also used to embed input mel spectrograms.

After autoregressive decoding,
the predicted spectrogram is refined
by a convolutional network as shown in Figure~\ref{fig:feedforward} similar to \citet{shen2018natural}.
By assuming the reconstruction likelihood as a product of two isotropic Gaussian distributions, i.e., $\mathcal{N}(x_t;\hat{x}_t,I)\mathcal{N}(x_t;\hat{x}_t+\mathrm{conv}(\hat{x})_t,I)$,
the reconstruction term $\mathbb{E}_{z_t \sim q} \big[\log p(x_t | z_t, x_{<t}, y)]$ in Eq \eqref{eq:vlb} is the mean squared error (MSE):
\begin{align}
\mathrm{MSE}(x, \hat{x}) = \frac{1}{T}\sum_{t=1}^T \big[\|x_t-\hat{x}_t\|^2_2 +
\|x_t-(\mathrm{conv}(\hat{x})_t + \hat{x}_t)\|^2_2 \big],
\label{eq:mse}
\end{align}
where $\mathrm{conv}(\cdot)$ stands for convolutional network.

\subsection{Slowness penalty}
To promote diverse generations, prior work has considered penalizing similarly synthesized acoustic frames \citep{meng-etal-2025-autoregressive}.
We introduce a slowness penalty to penalize slow changes in the predicted frames $\hat{x}$,
\begin{equation}
    \mathcal{L}_{\text{slow}} = -\frac{1}{T-1}\sum_{t=1}^{T-1} \|\hat{x}_t - \hat{x}_{t+1} \|^2.
    \label{eq:slow}
\end{equation}

%% file: sections/related.tex
\section{Related work}
We contrast MELD to prior work on autoregressive speech modeling in this section.
The proposed approach primarily differs from others in both how it derives input representations without relying on a pre-trained encoder or discretization step, and how the next prediction is sampled.

\textbf{Codec-based LMs.}
Recent advancements using discrete codecs derived from residual vector quantization (RVQ) have demonstrated success in zero-shot TTS \citep{chen2025neural,chen2024vall}.
The proposed approach does not apply RVQ to quantize mel-frames, since there are complications of using RVQ codecs.
In RVQ, each frame consists of multiple levels of codes that follow a fixed hierarchical order \citep{zeghidour2021soundstream,kumar2023high}.
Previous approaches predict this hierarchy using non-autoregressive models \citep{chen2025neural} or rearranging codes at each level with a delay pattern \citep{kharitonov-etal-2022-text,copet2023simple,defossez2024moshi}.
Our approach does not have such complications.
Moreover, codec-based LMs predict $N$ codebooks per time step. This increases output activation memory by a factor of $N$, resulting in higher GPU memory usage than MELD, which predicts a single latent token. 
The limitation is also noted in \citet{chou2025flow}. 

\textbf{Discretized mel spectrograms.}
To apply mel spectrograms for next-token prediction, \texttt{dMel} \citep{bai2024dmel} discretizes mel spectrograms with finite scaler quantization \citep{mentzer2023finite}.
We show that discretizing mel spectrograms before autoregressive modeling of speech is not necessary.
In fact, such discretization limits the performance of models on STT and joint TTS-STT modeling, as we will demonstrate in Sec \ref{sec:stt} and \ref{sec:ttsstt}.

\textbf{MELLE.} MELLE \citep{meng-etal-2025-autoregressive} is a mel-spectrogram-based model that draws samples from a single Gaussian autoregressively. 
In contrast, we predict the next frame by sampling over a categorical distribution mapped to $K$ isotropic Gaussians in a GMM (Sec \ref{sec:soft-vq}). We will show that this discrete latent space effectively suppresses stretched silence in synthesized speech in Sec \ref{exp:tts}.
In addition, the proposed objective forms a variational lower bound, 
but the objective of MELLE is designed in an ad-hoc manner (Appendix~\ref{app:melle}).
MELD does not need a stop predictor and can be easily extended to STT task. 
MELLE, however, focuses on speech generation and relies on voice activity detection (VAD) to avoid infinite silence generation.
We do not depend on such pre-processing step.

\textbf{End-to-end speech language modeling.} 
Although VoxCPM \citep{zhou2025voxcpm} is characterized as an ``end-to-end'' speech language model, it remains dependent on a pre-trained VAE to extract input representations.
In contrast, 
MELD directly models mel spectrograms without disconnected gradient flow, the encoder $g_{\text{mel}}$ is jointly optimized with the autoregressive model.

%% file: sections/exp.tex
\section{Experiments}
We conduct experiments to study whether discrete latent variables improve autoregressive mel spectrogram modeling (Sec~\ref{exp:tts},~\ref{sec:exp-latent}). 
We also study if optimizing directly over mel spectrograms improves STT compared to two-stage approaches (Sec~\ref{sec:stt}) and whether the strength of joint optimization transfers effectively to joint TTS-STT modeling (Sec~\ref{sec:ttsstt}).

\subsection{Experimental Settings}

We train all autoregressive models on the 960-hour subset of LibriSpeech \citep{panayotov2015librispeech}, referred to as LS960.
If not otherwise mentioned, we train all models using 12-layer Transformers with 16 attention heads, an embedding dimension (i.e., $d_{\text{model}}$) of 1024, a feed-forward network dimension of 4096, and a dropout rate of 0.2. 
The resulting model parameters are roughly 200M. We include more training details in Appendix~\ref{app:config}.

The choice of text tokens is critical to extend speech language models from TTS to STT. 
Unlike prior speech language models on TTS that heavily trained on phonemic transcriptions \citep{chen2025neural,chen2024vall,song2025ella,meng-etal-2025-autoregressive} on LS960, we adopt BPE tokenization with a vocabulary size $|\mathcal{V}_{\text{text}}|$ of 4096 for both TTS and STT modeling.
We compare two types of speech representations for speech language models: discrete speech codecs, due to their prominence in speech language modeling, and mel spectrograms.

\textbf{MELD implementation details.} Log mel spectrogram configurations are detailed in Appendix~\ref{app:mel}.
We encode mel spectrograms and codevectors with $g_{\text{mel}}$, a 3-layer MLP encoder with a network dimension of 1024.
Each layer is followed by a $\mathrm{GELU}$ activation function then a dropout of 0.5.
We follow the common practice, applying test-time dropout in $g_{\text{mel}}$ during the inference of zero-shot TTS \citep{wang2017tacotron,meng-etal-2025-autoregressive}.\footnote{
We find this necessary to mitigate the training-inference mismatch.}

We set the codebook size $K$ to 8192 and initialize it codebook with k-means++. The codebook is held fixed after training with $k$-means. 
The convolutional layers are based on the postnet of Tacotron 2 \citep{shen2018natural}, 
detailed in Appendix~\ref{app:conv}.
Mel spectrograms are converted into waveforms using a HiFi-GAN vocoder \citep{kong2020hifi} pre-trained on 585-hour LibriTTS.\footnote{\href{https://huggingface.co/mechanicalsea/speecht5-tts.}{https://huggingface.co/mechanicalsea/speecht5-tts.}}
Since our models are trained on normalized mel spectrograms,
the predicted frames are rescaled before the vocoder.

\begin{table*}[!ht]
\centering
\small
\caption{Summary of models for zero-shot TTS task w.r.t. their losses, the type of speech input and output, and the sampling methods, where $\mathcal{L}_{\text{stop}}$ is the binary cross-entropy loss of the stop predictor \citep{wang2017tacotron}.
}
\begin{tabular}{llllllll}
\toprule
\textbf{Model} & \textbf{Main Loss} &\textbf{Aux. Loss} &\textbf{Text} & \textbf{Speech} & \textbf{Vocoder} &\textbf{Sampling}  \\
\midrule
Codec-LM        & cross entropy &- &BPE & DAC code & DAC decoder & Top-$k$ / Top-$p$ \\
Mel-LM          & MSE \eqref{eq:mse} $+\mathcal{L}_{\text{stop}}$ &$\mathcal{L}_{\text{slow}}$ \eqref{eq:slow}&BPE & Mel   &HiFi-GAN      & No sampling \\
MELLE           &MSE \eqref{eq:mse} + $\mathcal{L}_{\text{stop}}$ &$\mathcal{L}_{\text{KL}}+\mathcal{L}_{\text{slow}}$ \eqref{eq:slow}  &BPE & Mel    &HiFi-GAN      & Gaussian \\
 MELD (proposed) &$\mathcal{L}_{\text{VLB}}$ \eqref{eq:vlb} &$\mathcal{L}_{\text{slow}}$ \eqref{eq:slow} &BPE & Mel  &HiFi-GAN   & Top-$k$ / Top-$p$ \\
\bottomrule
\end{tabular}
\label{tab:model_comparison}
\end{table*}

\subsection{A summary of model variants}
\label{sec:variant}
We summarize main model variants we are going to compare in Table \ref{tab:model_comparison}.
We apply top-$p$ sampling with $p=0.9$ and top-$k$ sampling with $k=60$ to sample from discrete distributions.
Moreover, we impose a score of $-1$ to the codes if they were sampled in the previous top-$p$ candidates, denoted as repetition penalty. 

\textbf{Codec-based baselines.} 
We implement a variant of our model based on speech codecs,  labeled as Codec-LM in Table~\ref{tab:model_comparison}.
We train Codec-LM by extracting speech codes using a DAC encoder that consists of 12 codebooks with a codebook size of 1024 \citep{kumar2023high}.\footnote{\href{https://github.com/descriptinc/descript-audio-codec}{https://github.com/descriptinc/descript-audio-codec}} 
Codebook embeddings are randomly initialized.
The DAC encoder encodes waveforms at 16kHz to a sequence of 12-level codecs at 50 Hz after RVQ \citep{zeghidour2021soundstream}.
To capture the dependency of residuals,
we introduce a delay of one code to each level of the RVQ codecs \citep{kharitonov-etal-2022-text,copet2023simple,defossez2024moshi}.
The delay pattern introduces a total delay of $(12 - 1) \times 20$ ms.
We use 12 linear heads to predict 12 levels of RVQ codes each time step for TTS.
The generation is terminated if any head predicts $\texttt{<EOS>}$.
The predicted codes are passed to the DAC decoder to synthesize waveforms.
To perform STT task, another linear head is used to predict BPE tokens.

While the codec-based baselines and mel-spectrogram-based baselines rely on different vocoders, DAC is trained on speech from multiple corpora comprising over 1000 hours of data \citep{kumar2023high}. 

\textbf{Mel-spectrogram-based baselines.} 
Two variants based on mel spectrograms are implemented by holding most of the architectures fixed. 
First, we remove the latent space and consider a decoder-only variant of Tacotron 2 \citep{shen2018natural}, denoted as Mel-LM.
The model is trained with MSE loss \eqref{eq:mse}.
No sampling is used for this variant.
Second, we reproduce MELLE  \citep{meng-etal-2025-autoregressive} following their settings.
MELLE differs from Mel-LM in terms of an additional KL term that regularizes Transformer outputs to a Gaussian prior, and samples predictions from a Gaussian distribution.
Both Mel-LM and MELLE rely on another linear layer to predict when to terminate,
while in MELD, $\texttt{<EOS>}$ is part of the discrete code vocabulary.
We apply a weight of 0.2 for slowness penalty $\mathcal{L}_{\text{slow}
}$ to all mel-spectrogram-based approaches. 
We detail the re-production of MELLE in Appendix \ref{app:melle}.

\begin{table}[h]
\centering
\small
\caption{Model comparisons between codec-based approaches and ours. All models are trained on LS960. $\clubsuit$ denotes the VALL-E results quoted from \citet{song2025ella}, which is trained on Encodec codecs derived from 24kHz audio samples.}
\begin{tabular}{llllll}
\toprule
\textbf{Model} & \textbf{Text} & \textbf{Speech} & \textbf{Freq} & \textbf{WER$\downarrow$} & \textbf{SIM$\uparrow$} \\
\midrule
Ground truth    & -- & -- & --   & 2.2 / 1.6 & 0.925 \\
DAC    & -- & -- & 50.0   & 2.2 / 1.6 & 0.922 \\
HiFi-Gan    & -- & -- & 62.5  &2.2 / 1.6 & 0.903 \\
\midrule
$\text{VALL-E }^\clubsuit$   & Phn & $\text{Encodec}$  & 75.0 & $\,\,\,$-$\,\,\,$ / 5.0 & 0.868 \\

Codec-LM    & Phn & $\text{DAC}$ & 50.0   & 5.7 / 4.7 & 0.872 \\
Codec-LM    & BPE & $\text{DAC}$ & 50.0   & 5.3 / 4.8 & 0.864\\\midrule
MELD   & BPE & Mel  & 62.5 & 2.4 / 1.9 & 0.872 \\
MELD   & BPE & Mel  & 31.3 & 2.5 / 1.9 & 0.855 \\

\bottomrule
\end{tabular}
\label{tab:compare-codec}
\end{table}

\subsection{Zero-shot text-to-speech}
\label{exp:tts}
We begin with zero-shot TTS task.
Given the text transcription and the first 3 seconds of an utterance as the prompt, 
we ask the model to continue the utterance with the hope of preserving speaker characteristics.
We follow the evaluation protocol in \citet{chen2025neural,borsos2023audiolm}, using audio samples with the duration between 4 seconds and 10 seconds from the LibriSpeech test-clean subset (2.2 hours).
We repeat speech generation 3 times and report the averaged scores.

We evaluate the generated samples with subjective and objective metrics.
For subjective metrics, 
two mean opinion scores (MOS) are evaluated---Similarity MOS (SMOS) and Comparison MOS (CMOS) \citep{ITUT,cooper2024review}.
We detail the subjective assessment in Appendix~\ref{app:mos}.
Content fidelity is evaluated by passing synthesized speech through a Conformer-Transducer and a Whisper-large model\citep{radford2023robust} to compute word error rates (WERs) against the ground truth.
We also evaluate speaker similarity between the given speech prompt (the first 3 seconds) and the generated speech. 
We follow \citet{song2025ella}, computing the cosine similarity (SIM) over speaker embeddings extracted from a WavLM-finetuned model,\footnote{\href{https://huggingface.co/microsoft/wavlm-base-sv}{https://huggingface.co/microsoft/wavlm-base-sv}.}
in which values above 0.86 are recognized as the same speaker.
Importantly, we use it solely as a proxy for assessing speaker similarity.

\begin{table}[t]

\centering
\begin{minipage}[t]{0.45\linewidth}
\centering
\small
\addtolength{\tabcolsep}{-0.1em}
\caption{Model comparisons of mel-spectrogram-based approaches. Predicted mel spectrograms are converted to waveforms with the same HiFi-GAN. Note that MELLE results are based on our reproduction.}
\begin{tabular}{lllll}
\toprule
\textbf{Model} &\textbf{$\mathcal{L}_{\text{slow}}$} & \textbf{Freq} & \textbf{WER}$\downarrow$ & \textbf{SIM}$\uparrow$ \\
\midrule
Mel-LM    & \checkmark   &62.5 & 4.7 / 4.2 & 0.825 \\
MELLE     & \checkmark     & 62.5 & 4.8 / 4.2 & 0.826 \\\midrule
MELD  & \checkmark   & 62.5 & 2.4 / 1.9& 0.872 \\
MELD   &\xmark & 62.5 &   6.0 / 3.7 &0.862  \\
MELD   & \checkmark  & 31.3 & 2.5 / 1.9& 0.855 \\
\bottomrule
\end{tabular}
\label{tab:compare-mel}
\end{minipage}
\hspace{0.3cm}
\begin{minipage}[t]{0.42\linewidth}
\centering
\small
\caption{Subjective evaluation was carried out on 43 samples with 40 speakers from the LibriSpeech test-clean set. CMOS is computed w.r.t. MELD.}
\begin{tabular}{llll}
\toprule
& \textbf{SMOS}$\uparrow$ & \textbf{CMOS}$\uparrow$ \\
\midrule
Ground Truth  &$4.11\pm0.10$  &0.27 \\
Codec-LM   &$3.72\pm0.15$ &-0.31\\
MELD (joint)  &$3.81\pm0.12$ &-0.20 \\
MELD   &$3.89\pm0.06$ &0.0 \\
\bottomrule
\end{tabular}
\label{tab:mos}
\end{minipage}
\end{table}

\textbf{Our Codec-LM is on par with VALL-E.} We first compare our Codec-LM with VALL-E \citep{song2025ella}, both are trained on LS960. 
For a closer comparison, we provide a Codec-LM baseline with phoneme tokens extracted from \citet{g2pE2019}.
In Table~\ref{tab:compare-codec}, our Codec-LM is on par with VALL-E in both content and speaker evaluations, despite VALL-E's two Transformers and higher frequency codecs derived from Encodec \citep{kumar2023high}.
We observe only a small decrease in speaker similarity when switching from phoneme tokens to BPE tokens in our Codec-LM.

\textbf{Joint optimization over mel spectrograms vs. two-stage approaches.}
We then compare MELD with Codec-LM. The two models share the same sampling strategies (Table~\ref{tab:model_comparison}). In particular, we find the repetition penalty effectively reduces the occurrence of long pauses and silence.
MELD obtains up to 2.3\% lower WERs compared to codec-based approaches, while maintaining comparable speaker similarity.
The improvements over Codec-LM also reflect on subjective evaluation in Table~\ref{tab:mos}.
When we reduce the frequency to 31.3 by concatenating every 2 frames,
we observe a decrease in speaker similarity, showing the importance of frame rates to preserve speaker similarity.
We will use MELD at a frequency of 62.5 Hz as our main baseline to the remaining experiments.

\textbf{Sampling from a discrete latent space is preferred over a single Gaussian.}
Regarding other approaches based on mel spectrograms in Table~\ref{tab:compare-mel}, they fall behind in both WERs and speaker similarity.
Although Mel-LM and MELLE have better WERs than Codec-LM in Table~\ref{tab:compare-codec}, 
the similarity scores fall behind by 0.047.
We consistently observe long pauses in samples generated by Mel-LM and MELLE despite the application of a slowness penalty during training.
While MELLE samples the next frame from a Gaussian latent space \citep{meng-etal-2025-autoregressive},
the latent space at $t$ is constrained to the choice of the prior $\mathcal{N}(x_t, I)$, due to the nature of the reverse KL divergence \citep{kingma2019introduction}.
Consequently, when a Gaussian is predicted to represent silence, it becomes unlikely to sample anything other than silence.
In contrast, the discrete latent space includes codes corresponding to both silent and non-silent frames, enabling MELD to escape the silence loop.

\begin{table}[h]
\centering
\small
\caption{The effectiveness of the discrete latent space during inference stage. \textbf{Mins} shows the total duration of synthesized samples. \textbf{S / D / I} is the WER breakdown, representing substitution, deletion and insertion errors.}
\begin{tabular}{lllll}
\toprule
& \textbf{WER}$\downarrow$ & \textbf{SIM}$\uparrow$ &  \textbf{Mins} &  \textbf{S / D / I}\\
\midrule
Ground truth &2.2 / 1.6 &0.925 & 131.8 & 0 / 0 / 0 \\
 MELD     & 2.4 / 1.9& 0.872 & 129.3 & 330 / 157 / 63\\
$\,$w/o rep penalty   & 3.1 / 2.6 &0.869 & 137.4 & 330 / 300 / 65\\
$\,$w/o $z_t$    & 52.3 / 51.7 & 0.520 &$>$200 &\\
\bottomrule
\end{tabular}
\label{tab:ablation}
\end{table}

\subsection{Effectiveness of the discrete latent space}
\label{sec:exp-latent}
Our objective admits collapsed solutions, e.g., constant $z_t$. However, we empirically observe that the method does not converge to such cases. To verify this, we set $c_{z_t}$ to zero when generating $x_t$ in \eqref{eq:specdec}. 
As shown in Table~\ref{tab:ablation},
we observe a substantial degradation of both WERs and speaker similarity when information from the discrete latents is removed, 
indicating that the discrete samples are essential to infer the next frames.
Unlike sampling from a Gaussian \citep{meng-etal-2025-autoregressive}, we can enhance sample diversity through repetition penalty that discourages repeated generations.
Removing the repetition penalty increases WER by 0.7\% with a slight decrease of speaker similarity, while the synthesized samples are noticeably longer ($\sim$6 mins) than the ground truth, suggesting excessive silence generation (Table~\ref{tab:ablation}).
Repetition penalty also effectively reduces deletion errors, suggesting that it helps mitigate word omissions.
We also study the effects of codebook size on the discrete latent space in Appendix~\ref{app:cluster}.

\begin{table}[h!]
\centering
\caption{Performance of decoder-only Transformers on LibriSpeech dev/test sets using LS960 hours. $\clubsuit$ Numbers quoted from \citet{defossez2024moshi} and \citet{bai2024dmel}, respectively.}
\small
\begin{tabular}{llllllll}
\toprule
&&&\multicolumn{2}{c}{\textbf{dev}}&\multicolumn{2}{c}{\textbf{test}} \\
 & \textbf{Size} & \textbf{Hrs} & \textbf{clean} &\textbf{other} & \textbf{clean} &\textbf{other}\\
\midrule
Moshi $^\clubsuit$         & 7B   & 7M    &- & - &5.8 & - \\
\texttt{dMel} (ASR) $^\clubsuit$ & 258M   & 960   & 3.8 & 10.3 &4.2 & 10.4 \\\midrule
Codec-LM   & 200M & 960  & 6.1 &16.5 &6.4 & 16.4  \\
$\,$w/o codebook init & 200M & 960  &$>$100 &$>$100&$>$100 &$>$100\\\midrule
MELD          & 200M & 960  & 4.0 &9.8 &4.2 & 10.0   \\
$\,$w/o SpecAug & 200M & 960  & 4.3 &12.5 &4.5 & 12.5   \\
MELD & 260M & 960  &3.6  &9.0 &3.5 & 9.2   \\
\bottomrule
\end{tabular}
\label{tab:stt}
\end{table}

\subsection{Speech-to-text with joint optimization}
\label{sec:stt}

Prior work on mel spectrogram language modeling only considers zero-shot TTS task \citep{meng-etal-2025-autoregressive}.
We extend MELD to STT task by putting speech input ahead of the BPE tokens as shown in Figure~\ref{fig:lm} (right).
We use the same decoder-only architecture in TTS for STT task but with two changes.
First, we deactivate dropout in $g_{\text{mel}}$ since we find it leads to unstable STT training.
Second, we use SpecAugment \citep{park2019specaugment}, applying two masks on frequency subbands with maximum frequency bands of 30 and 10 masks at frame level with maximum consecutive frames of 50. 
The number of frames per mask at frame level is clipped by $0.1$ times the number of frames.
Note that our MELD baseline on STT is equivalent to adding a linear head to MELLE to predict BPE tokens.

\textbf{Discretized representations fail to preserve task-specific information.}
Table~\ref{tab:stt} reports decoder-only STT results using beam search (a beam size of 5). 
We first note that, 
initializing codebooks for codec-based STT is critical for better convergence and WERs as noted in \citet{dhawan2024codec}.
We therefore initialize and freeze the codebooks from a pre-trained TTS-only Codec-LM.
As shown in Table~\ref{tab:stt}, using mel spectrograms consistently improves over Codec-LM for up to 6\% in test other, 
showing the strength of joint optimization in our framework.
We further compare MELD with \texttt{dMel} (ASR) that is trained on discretized mel spectrograms with an 18-layer Transformer decoder and the same SpecAugment settings \citep{bai2024dmel}.
Our approach outperforms \texttt{dMel} on decoder-only ASR model in dev-other (0.5\%) and both test-other (0.4\%). The gap is more pronounced with a larger Transformer (260M).

\begin{table}[h!]
\centering
\small
\caption{Results of MELD on joint TTS and STT modeling (joint), where we present STT results on test-clean/other. We compare the joint model with models trained on the separate tasks (separate).}
\begin{tabular}{llllllll}
\toprule
&&\multicolumn{2}{c}{\textbf{TTS}}&\multicolumn{2}{c}{\textbf{STT}} \\ 
& \textbf{Model} & \textbf{WER$\downarrow$} &\textbf{SIM$\uparrow$} & \textbf{clean} &\textbf{other}\\
\midrule
Moshi           & joint    &- & - &5.8 & - \\
\texttt{dMel} (ASR)    & separate & - & - &4.2 & 10.4 \\
\texttt{dMel} (ASR-TTS)    & joint   & - & - &7.5 & 15.3 \\\midrule
Codec-LM           &separate  & 5.3 / 4.8 &0.864 &6.4 &16.4    \\
MELD           &separate  & 2.4 / 1.9 &0.872 &4.2 &10.0    \\
MELD           & joint  &2.8 / 2.2 &0.870 &4.9 &12.1\\
\bottomrule
\end{tabular}
\label{tab:joint}
\end{table}

\subsection{TTS-STT modeling with MELD}
\label{sec:ttsstt}
Finally, we explore the joint modeling of both STT and TTS tasks with a 12-layer Transformer decoder.
As shown in Figure~\ref{fig:lm}, we can perform TTS given a text prompt and a control token \texttt{<TTS>}, or STT given a speech prompt and a control token \texttt{<STT>}, using a single speech language model.
We use a linear layer after the Transformer to jointly predict the discrete latent variables and BPE tokens and \texttt{<EOS>}.
During TTS, we mask the BPE tokens in $\mathcal{V}$ to avoid text predictions. Likewise, when decoding BPE tokens in STT, we mask the latent tokens in $\mathcal{V}$.

\textbf{Joint optimization reduces the gap between TTS-STT modeling and task-specific modeling.}

To enable both tasks, we adopt a simple multitask training strategy.
We initialize the training with TTS task for 80k steps, which is sufficient for a TTS model to converge. 
Then we continue the training with both tasks by mixing speech-text sequences for two modes equally in a training batch.
Also, we use a dropout of 0.5 on TTS during training and inference, while deactivating the dropout in STT mode.
We apply SpecAugment to STT input stream with two masks on frame level and two on frequency subbands, since we find the original configuration too aggressive and leads to worse WERs.
Table~\ref{tab:joint} reports the comparisons of joint models (joint) and task-specific models (separate).
MELD reduces the WER of \texttt{dMel} (ASR-TTS) on STT by 3.2\% on test-other. It also improves Codec-LM trained on separate tasks in both objective (Table~\ref{tab:joint}) and subjective (Table~\ref{tab:mos}) metrics. 
The results show that MELD not only supports high-quality zero-shot TTS but it can also perform STT task.

%% file: sections/conclusion.tex
\section{Conclusion}
In this work, we introduce MELD that predicts mel spectrogram frames through discrete latent variables. 
MELD presents a jointly optimized framework capable of TTS, STT, and TTS-STT modeling within a single model. 
Our results show the effectiveness of joint optimization in MELD, which improves STT performance over codec-based and discretized mel spectrogram baselines while maintaining strong zero-shot TTS quality. 
These findings show that mel spectrograms, combined with discrete latent modeling, provide an effective alternative to two-stage speech language modeling.

%% file: sections/limitation.tex
\section{Limitations}
We discuss several limitations of the proposed framework.
First, we acknowledge that it is hard to have fully fair comparisons between codec-based and mel-spectrogram-based methods. 
Although all models share the same Transformer decoder architecture, they differ in how speech representations are mapped to waveforms, using a DAC decoder for codecs or a HiFi-GAN for mel spectrograms.

Second, although we present several advantages relative to MELLE such as sampling discrete latents, we are not able to fully reproduce the results reported in MELLE \citep{meng-etal-2025-autoregressive} with LS960, while closely following their training configuration. 
We offer several possible explanations when compared to MELLE. We also detail our reproduction in Appendix~\ref{app:melle}.
We acknowledge that certain implementation details may have been overlooked during reproduction. 
For example, we suspect that the use of a VAD pre-processing step may be critical for achieving higher-quality samples, but not emphasized in the original work.

Last, we mainly focus on speech language models that model the conditional distributions of output speech (for TTS) and text (for STT).
We acknowledge that there are other speech tasks of interest to the community, such as question answering and speech translation \citep{arora2025landscape}. 
An immediate next step is to explore the proposed model on more speech tasks.

%% file: sections/ethical.tex
\subsection{Ethical considerations}
This work is intended to advance fundamental research in speech language modeling using mel-spectrograms and discrete latent variables.
All experiments are conducted on LibriSpeech, a publicly available dataset, used under the license of CC by 4.0.
Speakers are anonymized using numeric IDs.
Their use complies with the original dataset license and intended research purposes.

Potential ethical risks associated with this work include the misuse of speech generation methods, such as voice cloning. 
While the proposed models can perform zero-shot TTS, they are trained only on read speech. 
However, we acknowledge that the models can be potentially adapted to unseen speakers for realistic speech generation.
It is important for such systems to incorporate protocols to ensure speaker approval, regarding what data will be collected, how it will be used.

%% file: sections/appendix.tex
\section{Appendix}

\subsection{Variational Lower Bound} 
\label{app:proof}
To derive the variational lower bound we require the following two factorizations
\begin{align}
    q(z|x,y)&=\prod_t q(z_t|x_t)\notag\\
    p(x, z|y) &= \prod_t p(x_t, z_t|x_{<t}, z_{<t}, y)\notag\\
    &=\prod_t p(x_t|x_{<t},z_t, y)p(z_t | x_{<t}, y),\notag
\end{align}
the first equation is the proposal distribution with frame-wise dependency, while last line is due to the same assumptions we make in Section~\ref{sec:dlv-lm}, $z_t \perp z_{<t} \mid x_{<t}, y$ and
$x_t \perp z_{<t} \mid z_t, x_{<t}, y$.

With these two factorizations,
\begin{align}
&-\log p(x|y) \notag\\
&=-\log\sum_{z_1,\dots,z_T} p(x,z| y) \notag\\
&=-\log\sum_{z_1,\dots,z_T} \prod_t q(z_t|x_t)\frac{p(x,z| y)}{\prod_t q(z_t|x_t)} \notag\\
&=\sum_{t=1}^T -\log\mathbb{E}_ {z_t \sim q} \left[\frac{p(x_t|x_{<t},z_t, y)p(z_t | x_{<t}, y)}{q(z_t|x_t)} \right]\notag\\
&\leq \sum_{t=1}^T -\mathbb{E}_ {z_t \sim q} \left[\log\frac{p(x_t|x_{<t}, z_t,y)p(z_t | x_{<t}, y)}{q(z_t|x_t)} \right]\notag\\
&=
\sum_{t=1}^T \Big[
\mathrm{KL}\big[q(z_t|x_t)\,\|\, p(z_t | x_{<t}, y)\big] 
-\mathbb{E}_{z_t \sim q}\big[\log p(x_t |z_t, x_{<t}, y)\big]
\Big], 
\end{align}
we obtain the lower bound by Jensen's inequality.
We arrive at the variational lower bound of the conditional log likelihood \eqref{eq:vlb}.

\subsection{Training configuration}
\label{app:config}
We train all models for 200k steps on 16 NVIDIA V100 (16GB) GPUs with the Adam optimizer, using a maximum of 50k frames per batch and a gradient clip of 10. The learning rate is linearly warmed up for 1k steps to $5 \times 10^{-4}$, held constant for 100k steps, and then linearly decayed over the final 100k steps.
We also experiment with longer training steps, and we do not see notable improvements.

\subsection{Mel spectrograms}
\label{app:mel}
In line with the mel spectrogram settings from the pre-trained HiFi-GAN \citep{kong2020hifi}, we extract 80-dimensional log mel spectrograms with a frequency of 62.5 Hz (a frame shift of 16 ms), 64 ms frame length, Hann windowing, 1024-point Fourier transform. We apply mel filterbanks with the frequency range of 80 Hz to 7600 Hz. Mel spectrograms are then normalized with global mean and variance computed from the training data.

\subsection{Codebook size}
\label{app:cluster}

As shown in Table~\ref{tab:clustersize}, MELD is not sensitive to the choice of codebook size. This is because the quantized frames are further combined with the Transformer output ($h_t$ in Figure~\ref{fig:feedforward}) during mel spectrogram reconstruction.

\begin{table}[h]
\centering
\small
\caption{The effect of codebook size evaluated on WER and SIM metrics.}
\begin{tabular}{llll}
\toprule
\textbf{Codebook size} & \textbf{WER}$\downarrow$ & \textbf{SIM}$\uparrow$ \\
\midrule
4096  &1.9  &0.872 \\
8192  &1.9   &0.872\\
16384  &1.9 &0.871 \\
\bottomrule
\end{tabular}
\label{tab:clustersize}
\end{table}

\subsection{Subjective evaluation}
\label{app:mos}
We use SMOS to evaluate the
speaker similarity between the 3-second speech prompt and
the generated speech. 
We use CMOS to evaluate the comparative naturalness of the synthesized speech relative to the synthesized speech from the proposed approach MELD.
We recruited listeners via Amazon Mechanical Turk, a crowdsourcing platform. Each
screen contains two subtasks that are evaluated by 5 different listeners.

For each screen, listeners are presented with two samples for comparison and one sample as a reference.
We prepare 43 tests across 40 speakers in Librispeech test-clean set.
The first task (SMOS) is to rate the similarity of candidates (A and B) to the reference. The scale and the labels is defined in Table~\ref{tab:smos}.
Participants are instructed as follows:
\begin{itemize}
    \setlength{\itemsep}{-2pt}
    \item Rate how similar each sample is to the reference (not between A and B).
    \item Use the first 3 seconds of A and B as the reference and judge the similarity based on voice, speaking style, emotion, and audio quality.
\end{itemize}
For the second task (CMOS), the rating scale is defined in Table~\ref{tab:cmos} based on the following instruction:
\begin{itemize}
    \setlength{\itemsep}{-2pt}
    \item Compare which candidate sounds more natural (human-like).
\end{itemize}
A total of 30 listeners participated, with each screen receiving 15–20 scores.

\begin{table}[t]
\centering
\begin{minipage}[t]{0.47\linewidth}
\vspace{0cm}
\centering
\caption{Similarity rating scale used in SMOS.}
\setlength{\tabcolsep}{6pt}
\renewcommand{\arraystretch}{1.0}
\begin{tabular}{cl}
\toprule
\textbf{Score} & \textbf{Description} \\
\midrule
5   & Extremely similar \\
4.5 &  \\
4   & Very similar \\
3.5 &  \\
3   & Moderately similar \\
2.5 &  \\
2   & Slightly similar \\
1.5 &  \\
1   & Not similar at all \\
\bottomrule
\end{tabular}
\label{tab:smos}
\end{minipage}
\begin{minipage}[t]{0.47\linewidth}
\vspace{0cm}
\centering
\setlength{\tabcolsep}{6pt}
\renewcommand{\arraystretch}{1.0}
\caption{CMOS rating scale for naturalness comparison.}
\begin{tabular}{c l}
\toprule
\textbf{Score} & \textbf{Description} \\
\midrule
3  & A is much more natural than B \\
2  & A is more natural than B \\
1  & A is slightly more natural than B \\
0  & They are about the same \\
-1 & B is slightly more natural than A \\
-2 & B is more natural than A \\
-3 & B is much more natural than A \\
\bottomrule
\end{tabular}
\label{tab:cmos}
\end{minipage}

\end{table}

\subsection{Postnet}
\label{app:conv}
The postnet consists of a stack of 3 convolutional layers each containing 512 filters with shape $5 \times 1$ with batch normalization, followed by a $\mathrm{tanh}$ activation on every except the final layer.
The receptive filed is therefore $5\times 16$ ms.

\subsection{MELLE reproduction}
\label{app:melle}

The original objective of MELLE  \citep{meng-etal-2025-autoregressive} consists of four terms, a regression Loss $\mathcal{L}_{\text{reg}}$, KL Divergence Loss $\mathcal{L}_{\text{KL}}$, stop prediction $\mathcal{L}_{\text{stop}}$, and Spectrogram Flux loss $\mathcal{L}_{\text{flux}}$.
The losses are combined with different weights.

The regression loss is related to the MSE loss in our objective \eqref{eq:mse} for reconstruction but with extra L1 terms.
Their reconstruction loss only depends on $z_t$, while we have additional dependency on $x_{<t}$ and $y$.
Note that $z_t$ is ``continuous'' in MELLE.
We experiment with L1 loss following the original work but we do not find notable improvements.
The KL loss, $\mathrm{KL}(p(z_t|x_{<t}, y)\|\mathcal{N}(x_t, I))$, is introduced to map the Transformer output to a single Gaussian $\mathcal{N}(x_t, I)$ (the prior).
We follow \citet{meng-etal-2025-autoregressive}, using a linear layer to predict the mean and log-variance of the
Gaussian distribution for variational inference and sampling \citep{Kingma2014}, with a weight of 0.1 for $\mathcal{L}_{\text{KL}
}$.
We vary the weighting from 0.1 to 0.5 in increments of 0.1, but we find no notable improvements.

We experiment with the flux loss in \citet{meng-etal-2025-autoregressive} by varying the flux loss weight from 0.1 to 0.6 in increments of 0.1, but it leads to unstable training.
We instead use the proposed slowness penalty with a weight of 0.2 and find the slowness penalty ``necessary'' for MELLE and Mel-LM to suppress prolonged silence generations. 
We adopt the convolutional layers described in Tacotron 2 \citep{shen2018natural} to refine predicted mel spectrograms. 
Though the exact architectures of the convolutional layers are not presented in \citet{meng-etal-2025-autoregressive}.
We also experience with extending training MELLE for up to 400k steps, while the proposed approach and Codec-LM converges in 100k steps.

We suspect the VAD step is critical in the original MELLE to prevent models from keep producing silence.
Unfortunately, the original work does not provide results for experiments conducted without VAD, nor does it offer comparisons to other codec-based methods trained on audio samples processed with VAD.
It is also unclear how they filter ``abnormal'' silence.
Given that prior codec-based TTS studies  do not appear to have this pre-processing dependency \citep{song2025ella,han2024vall,chen2024vall}, we do not adopt VAD.